\documentclass[12pt]{article}

\usepackage{amsmath,amssymb,graphicx} 

\newcommand{\beq}{\begin{eqnarray}}
\newcommand{\eeq}{\end{eqnarray}}

\newcommand{\centeron}[2]{{\setbox0=\hbox{#1}\setbox1=\hbox{#2}\ifdim
                           \wd1>\wd0\kern.5\wd1\kern-.5\wd0\fi \copy0
                           \kern-.5\wd0\kern-.5\wd1\copy1\ifdim\wd0>\wd1
                           \kern.5\wd0\kern-.5\wd1\fi}}
\newcommand{\ltap}{\>\centeron{\raise.35ex\hbox{$<$}}
                   {\lower.65ex\hbox{$\sim$}}\>}
\newcommand{\gtap}{\>\centeron{\raise.35ex\hbox{$>$}}
                   {\lower.65ex\hbox{$\sim$}}\>}
\newcommand{\gsim}{\mathrel{\gtap}}

\newcommand\ZZ{\hbox{\zfont Z\kern-.4emZ}}
\font\zfont = cmss10 

\renewcommand{\theequation}{\thesection.\arabic{equation}}

\begin{document}
\begin{titlepage}
\begin{flushright}
{\tt hep-ph/0211124} \\
\end{flushright}

\vspace*{0.8cm}
\begin{center}
\vspace*{0.5cm}
{\LARGE \bf Big Corrections from a Little Higgs} \\
\vspace*{1.5cm}

\mbox{\bf
{Csaba Cs\'aki}$^{a}$, {Jay Hubisz}$^{a}$,
{Graham D. Kribs}$^{b}$,}\\
\mbox{\bf {Patrick Meade}$^{a}$, {and John Terning}$^{c}$} \\

\vspace*{0.8cm}

$^{a}$ {\it Newman Laboratory of Elementary Particle Physics, \\
Cornell University, Ithaca, NY 14853} \\
\vspace*{0.1cm}
$^{b}$ {\it Department of Physics, University of Wisconsin, Madison, WI 53706} \\
\vspace*{0.1cm}
$^{c}$ {\it Theory Division T-8, Los Alamos National Laboratory, Los Alamos,
NM 87545} \\
\vspace*{0.8cm}
{\tt  csaki@mail.lns.cornell.edu, hubisz@mail.lns.cornell.edu,
kribs@physics.wisc.edu, meade@mail.lns.cornell.edu, terning@lanl.gov}
\end{center}

\vspace*{1cm}

\begin{abstract}
\vskip 3pt
\noindent
We calculate the tree-level expressions for the electroweak
precision observables in the $SU(5)/SO(5)$ littlest Higgs model.
The source for these corrections are the exchange of heavy gauge
bosons and a triplet Higgs VEV.
Weak isospin violating contributions are
present because there is no custodial $SU(2)$ global symmetry.
The bulk of these weak isospin violating corrections arise from
heavy gauge boson exchange while a smaller contribution
comes from the triplet Higgs VEV.
A global fit is performed to the experimental data and we find that
throughout the parameter space the symmetry breaking scale is bounded by
$f > 4$ TeV at 95\% C.L.  Stronger
bounds on $f$ are found for generic choices
of the high energy gauge couplings.  We find that even in the best
case scenario one would need fine tuning of less than a percent
to get a Higgs mass as light as 200 GeV.
\end{abstract}

\end{titlepage}

\newpage


\section{Introduction}
\label{sec:intro}
\setcounter{equation}{0}
\setcounter{footnote}{0}

Recently there has been excitement generated by the
revival~\cite{little1,littlest,littlestmoose,Witek,littlepheno,Sekhar}
of the idea that the Higgs is a pseudo-Goldstone boson
\cite{HiggsPseudo,KaplanGeorgi,KaplanGeorgiSU2}.
New models have been generated
that provide
plausible realizations of this scenario with the feature that
they cancel all
quadratically
divergent contributions to the Higgs mass at one-loop. For a recent
review see~\cite{Martin}. These
``little'' Higgs models were
originally motivated~\cite{little1} by extra dimensional theories where the
Higgs is an extra component of the gauge fields~\cite{Manton,extra1,extra2,
extra3,extra4,extra5}, however the simplest little Higgs
models~\cite{littlest,littlestmoose,Witek} do not retain
any resemblance to extra dimensional theories.
At first glance these models seem
to allow a cutoff as large as 10 TeV without more than 10\% fine tuning in
the Higgs mass squared.  In particular a littlest Higgs model has
been proposed which is based on breaking an $SU(5)$ symmetry down to
$SO(5)$. When two $SU(2)$ subgroups (as well as two non-orthogonal
$U(1)$'s) of the $SU(5)$ are gauged they
are broken down to a diagonal $SU(2)_L\times U(1)_Y$ which can be
identified with
the electroweak interaction gauge group of the standard model (SM).
The main idea
behind the little Higgs models is that there are enough symmetries
in the theory that the simultaneous introduction
of two separate symmetry breaking terms is needed to force the Higgs to be a
pseudo-Goldstone boson
rather than an exact Goldstone boson. For example in the gauge sector of
the $SU(5)/SO(5)$ model
both $SU(2)$ gauge interactions are required to give a mass term to
the Higgs, thus the
quadratically divergent mass contributions must be proportional to
powers of both $SU(2)$ gauge couplings and hence can only appear at
two-loop order.

Since the weak interaction gauge group is a mixture of two
different gauge groups there are a variety of corrections to the
predictions for electroweak observables.  This type of mixing
correction is well known from previous extensions of the SM such
as the ununified model \cite{Georgiununified} and the $SU(3)$
electroweak model \cite{DimKap}. Current precision electroweak
data place constraints on the masses of the heavy analogues of the
$W$ and $Z$ to be of order 2-10 TeV depending on the model
\cite{CST,CEKT}. However in the little Higgs models the
corrections are more dangerous  since a ``custodial'' $SU(2)$
symmetry is not automatically enforced as it is in the SM or in
other words weak isospin is violated. (The importance of custodial
$SU(2)$ was recently emphasized in~\cite{Sekhar}.) These weak
isospin violating effects come both from heavy gauge boson
exchange and to a lesser extent from the presence of a triplet
Higgs VEV.

In the little Higgs
models raising the masses of the heavy gauge bosons means that log
divergent terms become large and hence that the fine-tuning becomes
more severe.
A similar effect happens with the masses of the fermions that cancel the
(even larger) top loop, which can make the situation even worse than one
might have expected.
For sufficiently large masses the large fine-tuning needed means that the
model fails to address the original motivation that inspired it.
In general there are two ways to try to satisfy the precision electroweak
bounds in these models. First one can take one of the two $SU(2)$
gauge couplings
to be large which reduces the direct coupling of quarks and leptons to the
heavy gauge boson; we will see that this approach turns out to be
unfavorable since it maximizes the mass mixing between heavy and light
gauge bosons which thus requires that the scale where the two gauge
groups break to $SU(2)_L$ to be large. This is problematic since it
raises the heavy
gauge boson and fermion masses and thus increases the fine-tuning as
described above.
The second approach is to tune the two  couplings (especially the
 two $U(1)$ couplings)
to be equal, then the mass mixing effects vanish and the bounds on
the high breaking scale are driven by the weak-isospin breaking
effects in four-fermion interactions,  which again result in a
strong bound on the breaking scale. In fact the correction to the
weak isospin violating $\Delta \rho_*$ parameter is
independent of the choice of the gauge couplings and  can be
brought to an acceptably small value only by raising the symmetry
breaking scale $f$.

In this paper we calculate the corrections to electroweak
observables in the littlest Higgs model.  We restrict ourselves to
tree-level effects. We then perform a global fit to the precision
electroweak data which allows us to quantify the bounds on the
masses of the heavy gauge bosons and fermions and thus also
quantify the required amount of fine-tuning. To illustrate the
importance of weak isospin breaking we perform fits with and
without the $SU(2)_L$ triplet\footnote{The appearance of the Higgs
triplet is not an essential part of little Higgs models, for
example the $SU(6)/Sp(6)$ model considered in~\cite{Witek} does
not have such a particle in its spectrum, instead it has two Higgs
doublets and an extra singlet.} VEV. We find that artificially
setting the triplet VEV to zero does not significantly improve the situation,
since there is still isospin breaking from heavy gauge boson
exchange.

\section{The Littlest Higgs Model at Tree-Level}
\label{spectrum-sec}
\setcounter{equation}{0}

We consider the little Higgs model based on the
non-linear $\sigma$ model describing
an \break
$SU(5)/SO(5)$ symmetry breaking~\cite{littlest}.
This symmetry breaking can be thought of as
originating from
a VEV of a symmetric tensor of the $SU(5)$ global symmetry. A convenient basis
for this breaking is characterized by the direction $\Sigma_0$ given by
\begin{equation}
\Sigma_0 =\left( \begin{array}{ccccc} &&&1& \\&&&&1\\ &&1\\1\\&1 \end{array}
\right).
\end{equation}
The Goldstone fluctuations are then described by the pion fields $\Pi =
\pi^a X^a$, where the $X^a$ are the broken generators of $SU(5)$. The
non-linear sigma model field is then
\begin{equation}
\Sigma (x) = e^{i\Pi/f} \Sigma_0 e^{i \Pi^T/f}=e^{2i\Pi/f} \Sigma_0.
\end{equation}
where $f$ is the scale of the VEV that accomplishes the breaking.
An
$[SU(2)\times U(1)]^2$ subgroup\footnote{Note that the two $U(1)$
generators
are not orthogonal and thus may have kinetic mixing terms, which will
imply additional corrections to electroweak observables.  Such kinetic
mixing terms are not generated at one-loop in the effective theory, but
may be generated by physics above the cut-off if there are heavy
particles charged under both U(1) groups.   Here we will assume
these effects are absent.}  of
the $SU(5)$ global symmetry
is gauged, where the generators of the gauged symmetries are given by
\begin{eqnarray}
&Q_1^a=\left( \begin{array}{ccc} \sigma^a/2 &0 & 0 \\
0 & 0 & 0\\ 0 & 0 & 0
\end{array}\right), \ \ \ &Y_1=
{\rm diag}(-3,-3,2,2,2)/10\nonumber \\
&Q_2^a=\left( \begin{array}{ccc} 0 & 0 & 0\\
0 & 0 & 0 \\
0 &0&-\sigma^{a*}/2\end{array} \right),
& Y_2={\rm diag}(-2,-2,-2,3,3)/10~,
\end{eqnarray}
where $\sigma^a$ are the Pauli $\sigma$ matrices. The $Q^a$'s are $5 \times 5$
matrices written in terms of $2 \times 2$, 1, and $2 \times 2$ blocks.
The Goldstone boson matrix $\Pi$,
in terms of the uneaten fields, is then given by
\begin{equation}
\Pi = \left( \begin{array}{ccc}
0 & \frac{H^\dagger}{\sqrt{2}} & \phi^\dagger
\\ \frac{H}{\sqrt{2}}& 0 & \frac{H^*}{\sqrt{2}}\\ \phi
&\frac{H^T}{\sqrt{2}} & 0
\end{array}\right),
\end{equation}
where $H$ is the little Higgs doublet $(h^0,h^+)$ and $\phi$ is a
complex triplet
Higgs, forming a symmetric tensor $\phi_{ij}$.
This triplet should have a very small expectation value
($\cal{O}$(GeV)) in order
to not give too large a contribution to the $T$ parameter.

We will write the gauge couplings of the $SU(2)$'s as $g_1$ and  $g_2$,
and similarly for the $U(1)$'s: $g_1^\prime$  and $g_2^\prime$.
We can assume that the quarks and leptons have their usual quantum numbers
under $SU(2)_L$
$\times$ U(1)$_Y$, but they are assigned under the first $SU(2) \times
U(1)$
gauge groups.

The kinetic energy term of the non-linear $\sigma$ model is \beq
\frac{f^2}{8} {\rm Tr} D_\mu \Sigma (D^\mu \Sigma)^\dagger \eeq
where \beq D_\mu \Sigma = \partial_\mu \Sigma - i \sum_j \left[
g_j W_j^a (Q_j^a \Sigma + \Sigma Q_j^{aT} ) + g_j^\prime B_j( Y_j
\Sigma + \Sigma Y_j)\right]~. \eeq

Thus at the scale of symmetry breaking $f$
(neglecting for the moment the Higgs VEV)
the gauge bosons of the four groups mix to form the
the light electroweak gauge bosons and heavy partners.
In the ($W_1^a$, $W_2^a$) basis (for $a=1,2,3$)
the mass matrix is:
\beq
\frac{f^2}{4} \left(\begin{array}{cc} g_1^2  &  -g_1 g_2  \\
-g_1 g_2  &g_2^2   \end{array}\right)
\eeq
Thus the light and heavy mass eigenstates  are:
\beq
W_L^a &=& s W_1^a + c W_2^a\\
W_H^a &=& -c W_1^a + s W_2^a
\eeq
with masses
\beq
M_{W_L} &=& 0 \\
M_{W_H} &=& \sqrt{g_1^2 + g_2^2} \frac {f}{2}
\eeq
where
\beq
s= \frac{g_2}{\sqrt{g_1^2+g_2^{2}}},\ \ \ \
c=\frac{g_1}{\sqrt{g_1^2+g_2^{2}}} ~.
\eeq

The $SU(2)$ singlet gauge bosons arise
from the U(1) gauge bosons $B_1$
and
the $B_2$.
The mass matrix in the ($B_1$,$B_2$) basis
at the high scale is
\beq
\frac{f^2}{20} \left(\begin{array}{cc} g_1^{\prime 2} &
   -g_1^\prime  g_2^\prime  \\
 - g_1^\prime  g_2^\prime
&g_2^{\prime 2} \end{array}\right)
\eeq
Thus the light and heavy mass eigenstates  are:
\beq
B_L &=& s^\prime B_1 + c^\prime B_2\\
B_H &=& -c^\prime B_1 + s^\prime B_2
\eeq
with masses
\beq
M_{B_L} &=& 0 \\
M_{B_H} &=& \sqrt{ g_1^{\prime 2} + g_2^{\prime 2}} \frac {f}{\sqrt{20}}
\eeq
where
\beq
s^\prime= \frac{g_2^\prime}{\sqrt{ g_1^{\prime 2}+g_2^{\prime 2}}},\ \ \ \
c^\prime=\frac{g_1^\prime}{\sqrt{g_1^{\prime ^2}+g_2^{\prime 2}}} ~.
\eeq

The effective gauge couplings of the $SU(2)_L$ $\times$ U(1)$_Y$ groups are:
\beq
g &=&  g_1 s\\
g^\prime &=& g_1^\prime s^\prime
~.
\eeq
Assuming that the first and second generation fermions transform only
under the $SU(2)_1\times U(1)_1$ gauge group,
the coupling of $W^a_H$ $(B_H)$ to quarks and leptons is
$- g_1 c$ $(- g_1^\prime  c^\prime)$.

\section{The Low-energy Effective Action}
\label{effective-sec}
\setcounter{equation}{0}
\setcounter{footnote}{0}

We now construct the effective theory below the mass scale of the
heavy gauge bosons. We obtain the gauge boson mass terms by
expanding the non-linear sigma field in the kinetic term to
quadratic order in $\Pi$.  At first glance, a quartic expansion is
necessary, however, the contributions coming
from higher powers of the expansion can be all absorbed by the shift of the
bare Higgs VEV $v$. Therefore even though the higher powers could contribute
to expressions in terms of bare parameters at the same order as the leading
piece, their effects will disappear from the final expressions in terms of
physical input parameters.  Also, all little Higgs physics which lead
to cancellations of quadratic divergences are captured at second order.

Integrating out $W^a_H$ and $B_H$ induces  additional
operators in the effective theory. These operators modify the usual
relations between the standard model parameters, and therefore their
coefficients can be constrained from electroweak precision measurements.
There are three types of operators that will be relevant for us: four-fermion
interactions,
corrections of the coupling of the $SU(2)_L \times U(1)_Y$ gauge
bosons to their corresponding currents, and operators that are quadratic
in the $SU(2)_L \times U(1)_Y$ gauge fields.
For simplicity we will work in a unitary gauge and
only keep track of the $h \equiv {\rm Re}\ h^0$ component of the Higgs field.

Exchanges of the  heavy $W^a_H$ and $B_H$ gauge bosons give the
following operators which are quadratic in the light gauge fields:
\begin{eqnarray}
{\cal L}_{2 {\rm mix}}&=&
-\frac{g^2  (s^2-c^2)^2 }{8 f^2} W^{a \mu}_L W^{a}_{\mu L}
h^{4}
- \frac{5 g^2  (s^{\prime 2}-c^{\prime 2})^2} {8 f^2} W^{3 \mu}_L
W^{3}_{\mu L} h^{4} \nonumber \\ &&
- \frac{g^{\prime 2}  (s^2-c^2)^2} {8 f^2} B^{\mu}_L B_{\mu L}  h^{4}
- \frac{5 g^{\prime 2}  (s^{\prime 2}-c^{\prime 2})^2} {8 f^2} B^{\mu}_L
B_{\mu L}  h^{4}
\\&&
+ \frac{g g^{\prime}  (s^2-c^2)^2} {4 f^2} W^{3\mu}_L B_{\mu L}  h^{4}
\label{operators1}
\end{eqnarray}
For example, the first term arises in the following way. The kinetic term of
the little Higgs field contains the coupling
\beq
{\cal L}_{\tilde{W}^2 h^{2}}=
\frac{g_1 g_2}{4}  W^a_{1 \mu} W_2^{a \mu} h^{2}~.
\eeq
Expressing $W_1$ and $W_2$ in terms of $W_L$ and $W_H$ we obtain a
coupling
between the heavy and light gauge bosons of the form
\beq
{\cal L}_{W_L W_H h^{2}}=-\frac{g_1 g_2 (g_1^2-
g_2^2)}{4(g_1^2+g_2^2)}  W^a_{\mu L}W^{a \mu}_H {h}^2~.
\eeq
The first term in ${\cal L}_{2 {\rm mix}}$ then arises by integrating out
the
heavy gauge boson $W^{a \mu}_H$ by taking its equation of
motion and expressing it in terms of the light fields.

In addition there are terms in the effective theory that are quadratic in
light gauge fields and
quartic in Higgs fields due to the expansion of the non-linear sigma field as well as couplings to the $SU(2)_L$ triplet $\phi$:
\begin{eqnarray}
{\cal L}_{2 {\rm nl}\sigma}&=&
\frac{g^2}{4 f^2} W^{a \mu}_L W^{a}_{\mu L} h^{4}
+ \frac{g^{\prime 2} }
{4 f^2} B^{\mu}_L B_{\mu L} h^{4}
-\frac{g g^\prime}{2 f^2} B^{\mu}_L W^{3}_{\mu L} h^{4}~,  \nonumber \\
{\cal L}_{2 \phi}
&=& \frac{g^2}{2} W^{a \mu}_L W^{a}_{\mu L} \phi^{2}  +
\frac{g^2}{2} W^{3 \mu}_L W^{3}_{\mu L} \phi^{2}
+ g^{\prime 2}  B^{\mu}_L B_{\mu L} \phi^{2}
-2 g g^\prime B^{\mu}_L W^{3}_{\mu L} \phi^{2}~.
\label{operators2}
\end{eqnarray}
We will only keep terms to order $\phi^2$ since $\langle \phi \rangle$
is phenomenologically
required to be small and is parametrically of order $v^2/f$ so corrections of
order $h^4/(v^2 f^2)$ and $\phi^2/v^2$ are actually the same order in
the $v/f$ expansion.

The operators in (\ref{operators1}) and (\ref{operators2})
give corrections to the light
gauge boson masses after the Higgs  gets a VEV. Thus
after $h$ and $\phi$ get VEVs:
\beq
\langle h \rangle&=& \frac{v}{\sqrt{2}}~,\\
\langle \phi \rangle &=& v^\prime~,
\eeq
and including the effects of the higher dimension operators
(\ref{operators1},\ref{operators2}), we find
that the mass of the $W$ is
\beq
M_W^2 &=&  g^2  \frac{v^2}{4}\left(1
+\frac{  (s^4+6s^2 c^2+c^4) v^2}{ 4  f^2}
+4\frac{v^{\prime 2}}{v^2} \right)~.
\eeq
Similarly, the mass of the $Z$ is
\beq
M_Z^2 &=&(g^2 + g^{\prime 2})\frac{v^2}{4}
\left(1+\frac{ ( s^4 +6s^2 c^2+c^4)v^2}{4 f^2}-\frac{5(s^{\prime
2}-c^{\prime 2})^2 v^2}{4 f^2}
+ 8\frac{v^{\prime 2}}{v^2} \right)~.
\label{Zmass}
\eeq

In addition, exchanges of $W^a_H$ and $B_H$ give
corrections to the coupling of the $SU(2)_L\times U(1)_Y$ gauge
bosons to their corresponding currents and additional four-fermion operators:
\beq
&&{\cal L}_{\rm c}=
g W^a_{L \mu} J^{a \mu}\left(1+
  \frac{c^2(s^2-c^2) h^{2}}{ f^2}\right)
+  g^\prime B_{L \mu} J^\mu_{Y}
\left(1- \frac{5 c^{\prime 2} (s^{\prime 2}-c^{\prime 2})
h^{2}}{f^2}\right)
\nonumber \\
&&+
g  W^3_{L \mu}J^\mu_{Y} \frac{5 (s^{\prime 2}-c^{\prime 2})h^{2}}{f^2}
-  g^\prime B_{L \mu} J^{3 \mu}
 \frac{c^2(s^2-c^2) h^{2}}{f^2}
-J_\mu^aJ^{a\mu} \frac{2 c^4}{f^2}-J_\mu^Y  J^{Y\mu}\frac{10 c^{\prime 4}}{f^2}
\label{currents}
\eeq

Using this expression we can now evaluate the effective Fermi coupling
$G_F$ in this theory. The simplest way to obtain the answer for this is by
integrating out the $W_L$ bosons from the theory by adding the $W$ mass term
to (\ref{currents}). The expression we obtain for the effective four-fermion
operator is
\begin{equation}
-\frac{g^2}{2M_W^2} J^{+\mu}J^-_\mu
\left[1+\frac{c^2(s^2-c^2) v^2}{f^2}\right]-
J^{+\mu}J^-_\mu \frac{2 c^4}{f^2}=-2 \sqrt{2} G_F
J^{+\mu}J^-_\mu,
\end{equation}
where $J^\pm=\frac{1}{2}(J^1\pm iJ^2)$.
Plugging in the correction to the $W$ mass we obtain that $G_F$ in this model
is corrected by
\begin{equation}
\label{treeGF}
\frac{1}{G_F}= \sqrt{2}v^2\left(1+\frac{v^2}{4f^2}
+4\frac{v^{\prime 2}}{v^2}\right).
\end{equation}

Finally, to fix all SM parameters we need to identify the
photon and the neutral-current couplings from (\ref{currents}):
\begin{eqnarray}
&&{\cal L}_{\rm nc}=
e A_{\mu} J^\mu_{Q}
+ \frac{e}{s_W c_W}Z_\mu\left[ J^{3 \mu}
\left(1
+  \frac{c^2(s^2-c^2)h^{2}}{f^2}
+ \frac{5c^{\prime
2}(s^{\prime 2}-c^{\prime 2})h^{2}}{f^2}\right) \right.
\nonumber  \\ && \left. - J^{\mu}_Q\left(s_w^2 +
5  \frac{c'^2(s'^2-c'^2)h^{2}}{f^2}
\right)\right]
-(J^3-J_Q)^\mu  (J^3-J_Q)_{\mu}\frac{10 c^{\prime 4}}{f^2}-\frac{2c^4}{f^2}
J^{3 \mu}J_\mu^3~ . \label{neutral}
\end{eqnarray}
Here $e= gg'/\sqrt{g^2+g^{\prime 2}}$ as in the standard model,
thus there is no
correction to the expression of the electric charge $e$
compared to the SM. Also $s_W^2$ in the above expression is the SM value
of the tree-level weak mixing angle $s_W^2=g'^2/(g^2+g'^2)$.
Similarly to the evaluation of the
effective $G_F$ we can calculate the low-energy effective four-fermion
interactions from the neutral currents. The result we obtain is
\begin{eqnarray}
{\cal L}_{NC} =&& -\frac{g^2+g'^2}{2 M_Z^2} \left[
J_3-s_W^2 J_Q+\frac{v^2}{2f^2}
(c^2(s^2-c^2) J_3 -
5 c'^2 (s'^2-c'^2) J_Y)\right]^2
\nonumber \\ && -\frac{10c'^4}{f^2} J_Y^2
 -\frac{2c^4}{f^2} J_3^2.
\label{NC}
\end{eqnarray}
where $M_Z$ is the physical $Z$ mass given in (\ref{Zmass}).

\section{The Contributions to Electroweak Observables}
\label{observables-sec}
\setcounter{equation}{0}
\setcounter{footnote}{0}
To relate the model parameters to observables
we use $\alpha(M_z)$, $G_F$, and $M_Z$ as input parameters.
We then use the
standard definition of the weak mixing angle
$\sin \theta_0$ from the $Z$ pole \cite{Sformulas},
\beq
\label{s2}
\sin^2 \theta_0 \cos^2 \theta_0 &=& \frac{\pi \alpha(M_Z^2)}{\sqrt{2} G_F M_Z^2}~,\\
\sin^2 \theta_0&=&0.23105 \pm 0.00008~,
\eeq
where~\cite{ErlerLang}
$\alpha(M_Z^2)^{-1}=128.92\pm 0.03$ is the running SM fine-structure
constant evaluated at $M_Z$.
We can relate this measured value with the bare value $s_W^2$
in this class of models, by using the expressions
 \beq
s_0^2\equiv \sin^2  \theta_0 &=& s_W^2 +\delta s_W^2
= s_W^2 -\frac{s_W^2 c_W^2}{c_W^2-s_W^2}
\left[ \frac{\delta G_F}{G_F}+\frac{\delta M_Z^2}{M_Z^2} \right]
\nonumber \\ &=& s_W^2-
\frac{s_W^2 c_W^2}{c_W^2-s_W^2}\left[4\Delta'+
\Delta \left(-\frac{5}{4} +c^2(1-c^2)+5 c'^2(1-c'^2)\right)\right],
\label{rens2}
\eeq
where we have defined
\beq
\Delta \equiv \frac{v^2}{f^2}, \qquad
\Delta'\equiv \frac{v'^2}{v^2}~.
\eeq
Also, we have the simple result that the
running couplings defined by Kennedy and Lynn~\cite{Lynn}
which appear in $Z$-pole
asymmetries are the same as
the bare couplings:
\beq
s^2_*(M_Z^2)=s_W^2, \ \ \
e^2_*(M_Z^2)=e^2 ~.
\eeq

In order to compare to experiments, we can relate our corrections of
the neutral-current couplings to the generalized modifications of the
$Z$ couplings as defined by Burgess et al. \cite{Burgess},
\begin{equation}
{\cal L}=\frac{e}{s_Wc_W} \sum_i \bar{f}_i \gamma^\mu
\left( (g_L^{f,SM}+\delta \tilde{g}_L^{ff}) P_L +
(g_R^{f,SM}+\delta \tilde{g}_R^{ff}) P_R \right) f_i
Z_\mu,  \end{equation}
where $P_{L,R}$ are left and right projectors, 
\begin{equation}
g_L^{f,SM} = t_3^f - q^f s_W^2 \quad , \quad g_R^{f,SM} = -q^f s_W^2
\end{equation}
are the SM couplings expressed in terms of $s_W^2$ that gets the 
correction (\ref{rens2}), and the overall coupling becomes
\begin{equation}
\frac{1}{s_W c_W} = \frac{1}{s_0 c_0}
\left[ 1 - 2\Delta'
-\frac{\Delta}{2}\left(-\frac{5}{4} + c^2(1-c^2)
+ 5c'^2(1-c'^2)\right) \right] \; .
\end{equation}
 From (\ref{neutral}) we obtain that
\begin{equation}
\delta \tilde{g}^{ff}=\Delta
\left[\frac{5t_3^f}{2} (c'^2-2c'^4) +\frac{t_3^f}{2}(c^2-2c^4)-q^f
\frac{5}{2} (c'^2-2 c'^4)\right].
\end{equation}
For the individual couplings this implies
\begin{eqnarray}
\delta \tilde{g}_L^{uu}&=& \frac{\Delta}{12} \left[3c^2(1-2c^2)-5c'^2(1-2c'^2)
\right],
\qquad\>\> \delta \tilde{g}_R^{uu}=
  -\frac{5 \Delta}{3} c'^2 (1-2c'^2), \nonumber \\
\delta \tilde{g}_L^{dd}&=&
  \frac{\Delta}{12}\left[ -3c^2(1-2c^2)-5c'^2(1-2c'^2)\right],
\quad  \delta \tilde{g}_R^{dd}=
 \frac{5\Delta}{6}c'^2(1-2c'^2), \nonumber \\
\delta \tilde{g}_L^{ee}&=&
  -\frac{\Delta}{4}\left[c^2(1-2c^2)-5c'^2(1-2c'^2)\right],
\qquad \delta \tilde{g}_R^{ee}=
  -\frac{5\Delta}{2} c'^2(1-2c'^2), \nonumber \\
\delta \tilde{g}_L^{\nu\nu}&=&
  -\frac{\Delta}{4}\left[-c^2(1-2c^2)-5c'^2(1-2c'^2)\right],
\end{eqnarray}
where $\delta \tilde{g}^{\mu\mu}=\delta \tilde{g}^{\tau\tau}=
\delta \tilde{g}^{ee}$, and similarly
$\delta \tilde{g}^{tt}=\delta \tilde{g}^{cc}=\delta \tilde{g}^{uu}$,
$\delta \tilde{g}^{bb}=\delta \tilde{g}^{ss}=\delta \tilde{g}^{dd}$.

In order to calculate the corrections to the low-energy precision
observables for neutrino scattering and for atomic parity
violation we need to write the low-energy
effective neutral current interaction (\ref{NC}) in
the form
\begin{equation}
-\frac{4 G_F}{\sqrt{2}} \rho_* (J_3-s_*^2(0) J_Q)^2 +\alpha J_Q^2.
\end{equation}
Here $s_*^2(0)$ is the low-energy value of the effective Weinberg angle,
different from $s_*^2 (M_Z^2)$. The last term proportional to $\alpha$ will not contribute to any of the
low-energy processes we are constraining, therefore what one needs to do
is to express $\rho_*$ and $s_*^2$ in terms of our variables
$\Delta,\Delta' ,c,c'$.
We find the following expressions:
\begin{eqnarray}
&& \rho_* = 1-4\Delta' +\frac{5}{4} \Delta \nonumber \\
&& s_*^2(0)
= s_W^2-\frac{\Delta}{2}\left[ s_W^2 ( c^2+5 c'^2)-5 c'^2\right],
\label{rhostar}
\end{eqnarray}
where $s_W^2$ has to be expressed in terms of $s_0^2$ using (\ref{rens2}).
Note that $s_*^2(0) \ne s_*^2(M_Z)$ due to the corrections to four-fermion
interactions from heavy gauge boson exchange.
The low-energy observables can then be expressed using the relations
\begin{eqnarray}
&g_L^2= \rho_*^2\left[\frac{1}{2}-s_*^2(0)+\frac{5}{9} s_*^4(0)\right]
~,\ \ \ \ &
g_{eV}(\nu e \rightarrow \nu e) = 2\rho_*\left[s_*^2(0) - \frac{1}{4}\right]~,
\nonumber \\
&g_R^2=\rho_*^2 \frac{5}{9} s_*^4(0), &
g_{eA}(\nu e \rightarrow \nu e) = -\frac{\rho_*}{2}~, \nonumber \\
& Q_W(Z,N)=-\rho_* \left[ N-(1-4 s_*^2(0)) Z\right]~.
\end{eqnarray}
In the Appendix we calculate the shifts in the electroweak precision
observables in terms of the parameters $c,c',\Delta$  and $\Delta'$
defined above.

Until now we have treated $\Delta$ and $\Delta'$ as independent variables.
However the triplet VEV is not independent from the Higgs VEV, since it
arises from the same operators that give rise to the quartic scalar potential
of the little Higgs~\cite{littlest}.
The leading terms are the quadratically divergent
pieces in the Coleman-Weinberg (CW) potential (which by construction does
not contribute to the little Higgs mass) and their tree-level counter terms.
For example the gauge boson contribution to the CW potential is
\begin{equation}
\frac{\Lambda^2}{16 \pi^2} {\rm Tr} M_V^2 (\Sigma ),
\end{equation}
where $M_V^2$ is the gauge boson mass matrix in an arbitrary $\Sigma$
background. For example the first 3$\times$3 block of the gauge boson
mass matrix (corresponding to the first $SU(2)$ gauge group) is~\cite{littlest}
\begin{equation}
M_V^{2\ ab}=g_1^2 {\rm Tr} (Q_1^a\Sigma+\Sigma Q_1^{a\, T})(\Sigma^{\dagger}
Q_1^b+Q_1^{b\, T} \Sigma^\dagger ).
\end{equation}
Evaluating the full expression for the gauge boson contributions
results in a potential (to cubic order) of the form
\begin{equation}
V_{GB}= a f^2\left[ (g_2^2+g_2'^2) \left| \phi_{ij}+\frac{i}{4f}(
h_i h_j+ h_j h_i)\right|^2+ (g_1^2+g_1'^2) \left|
\phi_{ij}-\frac{i}{4f}( h_i h_j+ h_j h_i)\right|^2\right],
\label{CW1}
\end{equation}
where $a$ is a constant of order one determined by the relative
size of the tree-level and loop induced terms, and we have used
$\Lambda \sim 4\pi f$. Similarly, the fermion loops contribute
\begin{equation}
-a' \lambda_1^2 f^2\left| \phi_{ij}+\frac{i}{4f}( h_i h_j+
h_j h_i)\right|^2,
\label{CW2}
\end{equation}
where $\lambda_1$ (and $\lambda_2$) are the Yukawa couplings and mass
terms.

The Yukawa couplings for the light and heavy top quarks come from
expansion of the operator
\begin{equation}
\frac{1}{2} \lambda_1 f \epsilon_{i j k} \epsilon_{x y} \chi_i
\Sigma_{j x} \Sigma_{k y} u'^c_3
\end{equation}
where $\chi = (b_3\ t_3\ \tilde{t})$.  The resulting Lagrangian
(up to irrelevant phase redefinitions of the quarks) is given by
\begin{equation}
{\cal L}_{fermion}= \lambda_1 (\sqrt{2} q_3 h +f \tilde{t})u'^c_3 +\lambda_2 f \tilde{t}\tilde{t}^c.
\end{equation}
where $q_3 = (b_3\ t_3)$. Here $\tilde{t},\tilde{t}^c$ are the
extra vector-like color triplet fermions necessary to cancel the
quadratic divergences to the little Higgs mass from the top loops,
and the physical right-handed top is $(\lambda_2 u'^c_3-\lambda_1
\tilde{t}^c)/\sqrt{\lambda_1^2+\lambda_2^2}$ with a Yukawa coupling
$\lambda_t = \sqrt{2}
\lambda_1\lambda_2/\sqrt{\lambda_1^2+\lambda_2^2}$. The mass of
the heavy fermion is $\sqrt{\lambda_1^2+\lambda_2^2}f \equiv
\lambda_H f$.

Thus one can see from (\ref{CW1}) and (\ref{CW2}) that there must be a triplet
VEV of order $v' \sim v^2/f$ as we have assumed before. In terms of the
parameters $a$, $a'$,and $\lambda_1$ the triplet VEV is given by
\begin{equation}
v'=-\frac{i v^2}{4f} \frac{a(g_2^2+g_2'^2-g_1^2-g_1'^2)-
a'\lambda_1^2}{a(g_2^2+g_2'^2+g_1^2+g_1'^2)- a'\lambda_1^2}.
\label{tripletvev}
\end{equation}
However, since the terms (\ref{CW1}) and (\ref{CW2}) are also responsible
for the quartic scalar coupling of the little Higgs, one can eliminate the
parameters $a',\lambda_1$ from the expression for the triplet VEV. The
quartic scalar coupling is given by~\cite{littlest}
\begin{equation}
\label{lambda} \lambda
=\frac{[a(g_2^2+g_2'^2)-a'\lambda_1^2][a(g_1^2+g_1'^2)]}{a(g_2^2+g_2'^2+g_1^2+g_1'^2)-a'\lambda_1^2},
\end{equation}
and thus we obtain that
\begin{equation}
\Delta' \equiv \frac{|v'|^2}{v^2} = \frac{\Delta}{16} \left[
\frac{2 \lambda}{a(g_1^2+g_1'^2)}-1\right]^2~. \label{deltaprime}
\end{equation}
For a Higgs mass of order 200 GeV, $\lambda\approx 1/3$ (at
tree-level). The above formula can then be used for the fit with
reasonable values of the coefficient $a$.  There is one further
constraint on the parameters of the model.  In ref.
\cite{littlest} it was shown that in order to have a positive mass
squared for the triplet one must have: 
\beq
a(g_1^2+g_1'^2+g_2^2+g_2'^2)>a'\lambda_1^2 \label{aconstraint}
\eeq 
which is equivalent (using our previous constraint (\ref{lambda})) to
requiring
\beq
a > \frac{\lambda}{g_1^2+g_1'^2}~. \label{abound}
\eeq
If the triplet mass squared is negative it   implies that the
triplet gets a VEV of order $f$, which is impossible to reconcile
with electroweak data. This constraint will prove to be important
since it excludes the region of large triplet VEVs. From Eq.
(\ref{deltaprime}) we see that it enforces
\begin{equation}
\Delta' < \frac{\Delta}{16}~.
\label{deltaprimebound}
\end{equation}
Actually the bound on $a$ is more severe than Eq. (\ref{abound})
due to the non-observation of the triplet scalar at LEP, but
Eq. (\ref{abound}) will be sufficient for our purposes.

There will also be contributions to observables due to heavy quark
loop modifications of the light gauge boson propagators. For example, these
would result in contributions to the
$\rho_*$-parameter. To compute these loops, we diagonalize the top
quarks into a mass eigenbasis,
\begin{eqnarray}
t_3 = s_2 t_3^H + c_2 t_3^L \nonumber \\
\tilde{t} = c_2 t_3^H - s_2 t_3^L
\end{eqnarray}
with
\begin{equation}
s_2 = \frac{\lambda_1^2}{\lambda_1^2+ \lambda_2^2} \frac{ v }{ f }
\end{equation}
After expressing the interactions with the SM gauge bosons in
terms of heavy and light Dirac fermions, we compute all loop
corrections, and subtract off standard model contributions. We
find that for generic values of the Yukawa couplings $\lambda_1$
and $\lambda_2$, these corrections to $\rho$ are suppressed by
$\Delta$, as well as a $\frac{1}{16 \pi^2}$ loop factor. The
leading order term is given by:
\begin{equation}
\Delta \rho_{\mathrm{top}} = \frac{ N_c \Delta}{16 \pi^2} \left(
\frac{ \lambda_t^4 }{2 \lambda_H^2} \right) \log \left[ \frac{ 2
  \lambda_H^2 }{\lambda_t^2 \Delta} \right].
\end{equation}
This expression should be compared to the smaller of the
contributions we have included in our calculation, arising from
the triplet VEV $\Delta \rho_{triplet} = -4 \Delta' \sim \Delta
/4$ for large values of $a$. However $\lambda_H \geq \sqrt{2}$,
and since $\frac{\log 2 \lambda_H^2/\Delta}{\Lambda_H^2}$ has its
maximum for $\lambda_H =\sqrt{2}$, the leading piece of the top
contribution is at most $-\frac{3}{64 \pi^2}\log
(\frac{v^2}{4f^2}) \Delta$, which for $f\sim 4\;\mbox{TeV}$ is
about $0.032 \Delta$, almost an order of magnitude smaller \ than
the maximal contribution of the triplet vev. Thus it is well
justified to ignore these loop effects in the fits.

\section{Results and Interpretation}
\label{results-sec}
\setcounter{equation}{0}
\setcounter{footnote}{0}

\begin{figure}[t]
\centerline{\includegraphics[width=0.75\hsize]{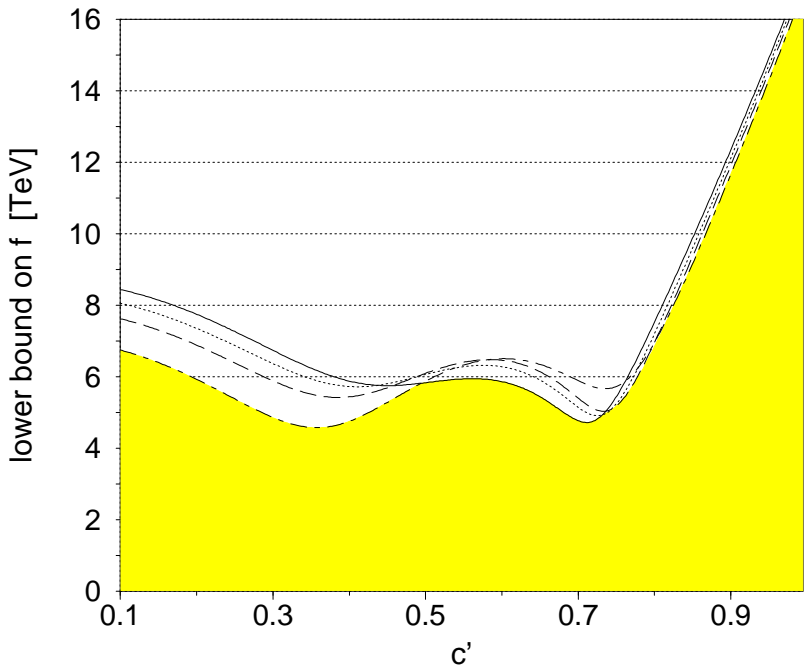}}

\caption{The region of parameters excluded to 95\% C.L.
is shown as a function of $c'$.  The region below the contours is excluded
to 95\% C.L. for $c$ equal to $0.1$ (solid), $0.5$ (dotted),
$0.7$ (dashed), $0.99$ (dot-dashed). The shaded region is excluded
for any choice of $c$.  }
\label{contour-fig}
\end{figure}

Since the parameter $a$ is expected to be ${\cal O}(1)$, we will consider
fixed values of $a$ in the range 0.1 - 2.
However to begin the discussion of our results we will artificially set
the triplet VEV to zero.
This not only makes the analysis and interpretation simpler it also
contains the essential physics that constrains generic little Higgs models.
We performed a three parameter global fit
(as described in \cite{RSfit}) to the 21 precision
electroweak observables given in Table~\ref{table}.
The best fit was found to be for $c \simeq 1$,
$c' \simeq 0.32$, and $f\approx8.9$
TeV\@, with a $\chi^2$ per degree
of freedom ($21-3=18$):
\begin{equation}
\frac{\chi^2_{\rm best}}{(\mbox{d.\ of f.})} \simeq 1.54
\end{equation}
that is slightly worse than the fit to the SM,
\begin{equation}
\frac{\chi^2_{\rm SM}}{(\mbox{d.\ of f.})} \simeq 1.38 \; .
\end{equation}
\begin{figure}[t]
\centerline{\includegraphics[width=0.75\hsize]{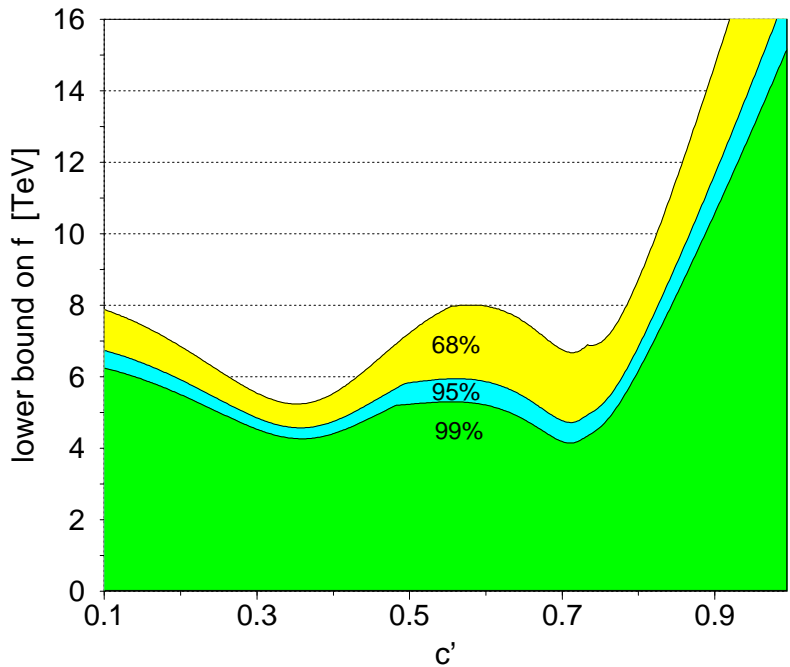}}

\caption{The region of parameters excluded to 68\%, 95\%, and 99\% C.L.
is shown as a function of $c'$.  The parameter $c$ was allowed
to vary between $0.1 < c < 0.995$ for each $c'$ to give the least
restrictive bound on $f$.  (See also Fig.~\ref{contour-fig}.)}
\label{limit-fig}
\end{figure}
First consider the region of parameters relevant to the model.
To ensure the high energy gauge couplings $g_{1,2},g'_{1,2}$ are
not strongly coupled, the angles $c = g/g_2$, $s = g/g_1$
$c' = g'/g'_2$, $s' = g'/g'_1$ cannot be too small.  We
conservatively allow for $c,s,c',s' > 0.1$, or equivalently
$0.1 < c,c' < 0.995$.  We allow $f$ to take on any value
(although for small enough $f$ there will be constraints from
direct production of $B_H$).  The general procedure we used is to
systematically step through values of $c$ and $c'$, finding the
lowest value of $f$ that leads to a shift in the $\chi^2$
corresponding to the 68\%, 95\%, and 99\% confidence level (C.L.).
For a three-parameter fit, this corresponds to a $\Delta \chi^2$
of about $3.5$, $7.8$, $11.3$ from the minimum, respectively.
Globally, for any choice of high energy gauge couplings, we find
\begin{equation}
f > (4.1, \, 4.6, \, 5.2) \; {\rm TeV} \quad {\rm at} \quad
(99\%, \, 95\%, 68\%) \; {\rm C.L.}
\end{equation}
We used $m_h = 115$ GeV, and verified that the bound is not lowered
for larger values of the Higgs mass.
Of course these bounds are saturated only for very specific values
of the gauge couplings.  The bound on $f$ is perhaps best illustrated as
a function of $c'$, which we do in Fig.~\ref{limit-fig}.  The shaded
area below the lines shows the region of parameter space excluded by
precision electroweak data.  Note that
we numerically found the value of $c$ that gave the
\emph{least restrictive} bound on $f$ for every $c'$.  For a specific choice
of $c$ the bound on $f$ can be stronger.
This is shown in Fig.~\ref{contour-fig}
where we show contours of the 95\% excluded region for fixed $c$
while $c'$ was allowed to vary.  This figure makes it clear that
the least restrictive bound is obtained for different values of
$c$ as $c'$ is varied.  The shaded region is identical to
the 95\% C.L. region shown in Fig.~\ref{limit-fig}, illustrating
how the exclusion regions in Fig.~\ref{limit-fig} were obtained.

\begin{figure}[t]
\centerline{
\includegraphics[width=0.55\hsize]{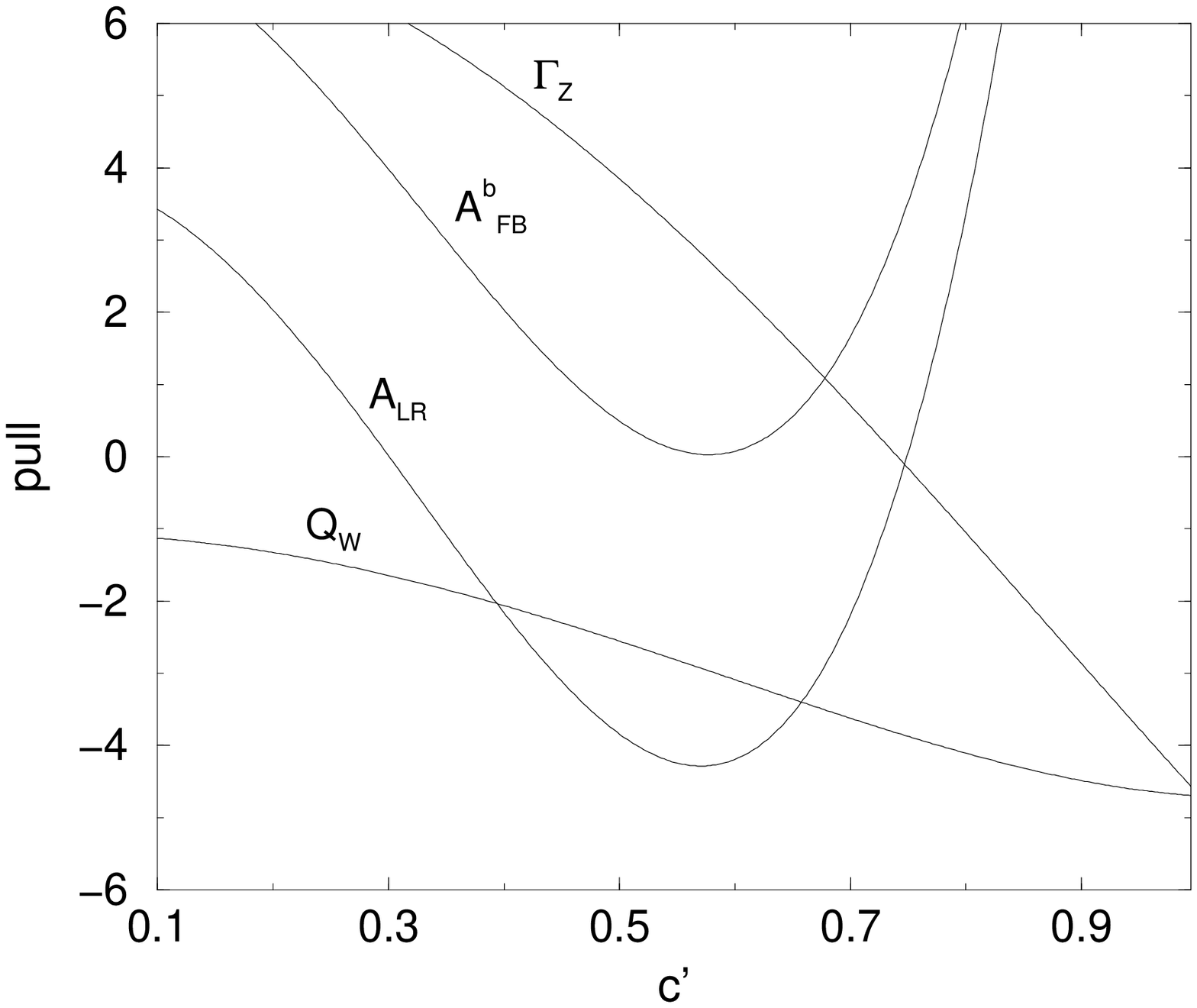}
\hfill
\includegraphics[width=0.55\hsize]{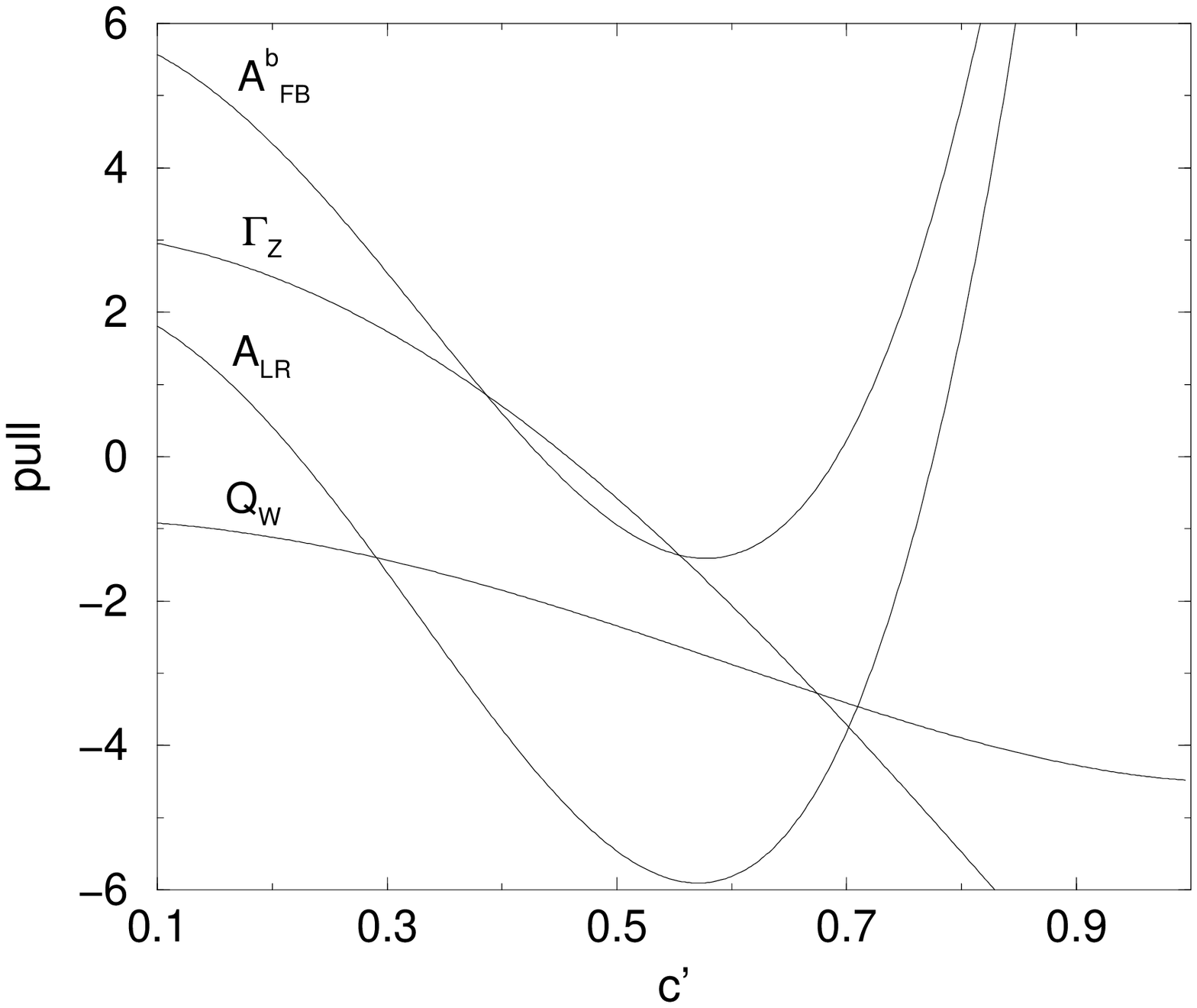}}
\caption{The difference between the predicted and the experimentally
measured values for four observables in standard deviations (the pull).
In both figures the decay constant was fixed to $f=4$ TeV\@.  The figure
on the left (right) has a fixed $c=0.1$ ($c=0.99$); other values
interpolate between these two figures.  Notice that the observable
that gives the largest contribution varies depending on $c'$.}
\label{observables-fig}
\end{figure}

While we fit to 21 observables, inevitably certain
observables are more sensitive to the new physics.
To gain some insight into the main
observables leading to the bounds shown in
Figs.~\ref{limit-fig}--\ref{contour-fig}, we show in
Fig.~\ref{observables-fig} the deviation of $\Gamma_Z$, $A^b_{FB}$,
$A_{LR}$, and $Q_W$ from the experimental value (the pull) as a
function of $c'$ for fixed $f = 4$ TeV and two choices of
$c = 0.1$ and $0.99$.  A set of parameters is typically ruled out
by the global fit once a single observable has a pull
greater than of order $\pm 4$.  The variation of $A^b_{FB}$ and
$A_{LR}$ also explains the appearance in
Figs.~\ref{limit-fig}--\ref{contour-fig} of a rise in the bound
on $f$ for small $c'$ (where $A^b_{FB}$ is important) large
$c'$ (where many observables are important) and the bump in the
middle (where $A_{LR}$ is important). Note that the region of large
$c^\prime$ corresponds to the $U(1)_1$ gauge coupling
(the gauge coupling of the quarks and leptons) getting strong.

\begin{figure}[t]
\centerline{\includegraphics[width=0.75\hsize]{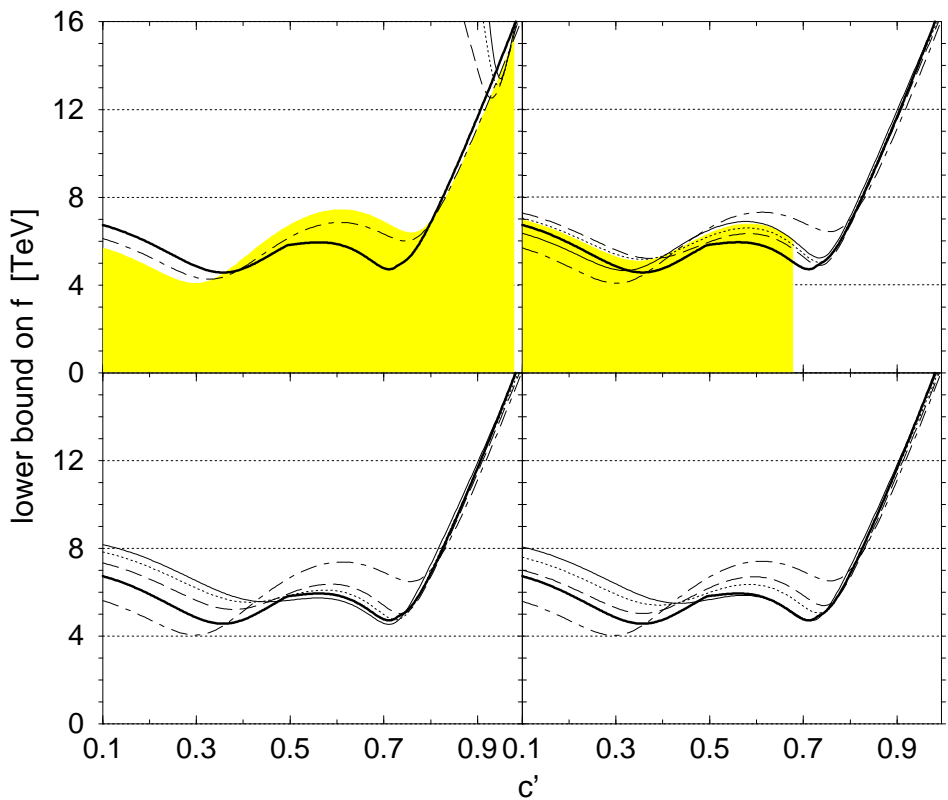}}
\caption{The region of parameters excluded to 95\% C.L.
is shown as a function of $c'$.  The four figures correspond
to $a=0.1,0.5,1,2$ corresponding to top-left, top-right, bottom-left, and
bottom-right. The region below the contours is excluded
to 95\% C.L. for $c$ equal to $0.1$ (solid), $0.5$ (dotted),
$0.7$ (dashed), $0.99$ (dot-dashed). The heavy solid line displays the bound
from Fig. 2. The shaded region corresponds to
the extension of the excluded  region obtained by requiring a positive triplet
mass squared. }
\label{C-fig}
\end{figure}

\begin{figure}[t]
\centerline{\includegraphics[width=0.75\hsize]{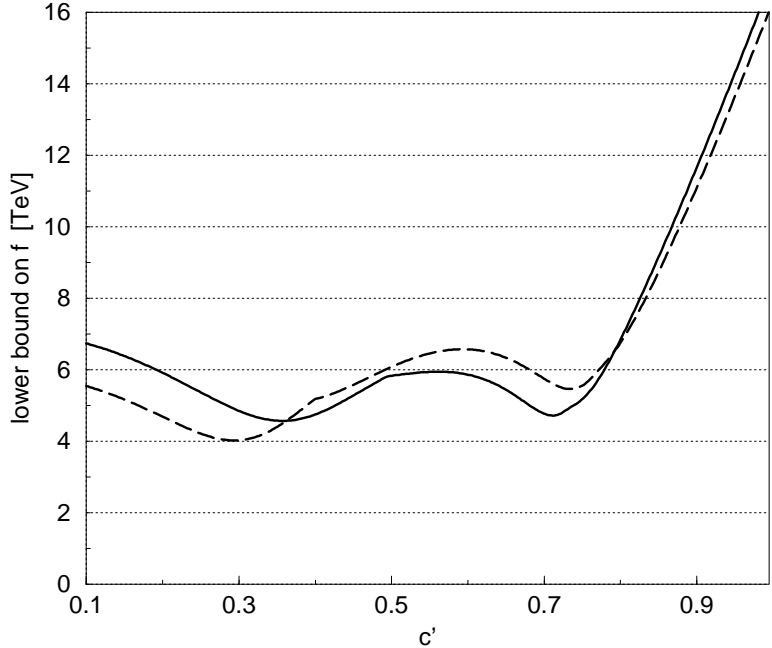}}
\caption{The region of parameters excluded to 95\% C.L.
is shown below the lines. The solid curve is for $\Delta^\prime=0$
as in Fig. 1, while the dashed curve is for $\Delta^\prime=\Delta/16$.
The weakest bound at $c^\prime=0.29$ corresponds to $a=0.025$
[using Eq.(\ref{deltaprime})]. }
\label{b-fig}
\end{figure}

We can now move on to discuss the fits done for reasonable values
of the parameter $a$. We have redone the fit with
$a=(0.1,0.5,1,2)$ and the results are displayed in Fig.~\ref{C-fig}. 
For generic couplings we see that the 95\% C.L. bound
is well above 4 TeV.  However, for certain choices of the
parameters the contribution from the triplet VEV can 
partially cancel against the gauge boson contributions
accidentally.  This happens near $c' \sim 0.3$ when the
triplet VEV, Eq.~(\ref{deltaprime}), is maximized which
occurs when either
\begin{equation}
\frac{\lambda}{a (g_1^2 + g_1'^2)} \sim 1 \quad {\rm or} \quad
\frac{\lambda}{a (g_1^2 + g_1'^2)} \ll 1 \; .
\end{equation}
The first region corresponds to very small $a \sim 0.025$ where
the triplet mass, Eq.~(\ref{aconstraint}), nearly vanishes.
The second region corresponds to larger values of $a \gsim 1$, 
as illustrated by the bottom two figures.  However, as pointed out 
in \cite{littlest} the triplet VEV is naturally expected to be small, 
and we see from Fig.~\ref{C-fig} that it is generically a sub-leading 
effect.  The weakest bound, $f > 4.0$ TeV, arises for $a=0.025$ 
or $a \gsim 1$.  Another way to obtain an approximate bound is to
redo the fit imposing a maximal triplet VEV, $\Delta^\prime = \Delta/16$, 
shown in Fig.~\ref{b-fig}. In general the real bound on $f$ could be stronger 
than what is found by this method, because the point corresponding to
this bound might be eliminated by imposing the constraint of a
positive triplet mass squared.  This method gives a bound similar to
that obtained from Figure \ref{C-fig} because at the lowest value of
$f$ the triplet mass squared is positive.  Similarly, one finds that
the lower bound on the heavy gauge boson masses are:
$M_{B_H}>650$ GeV and $M_{W_H}> 2.7$ TeV.


Given the above bound on the scale $f$ we can quantify the amount
of fine-tuning in the model using the fine-tuning measure proposed
in \cite{littlest}. The contribution to the Higgs mass squared
from the heavy partner of the top (with mass $m^\prime$) is
 \beq -
\frac{3 \lambda_t^2}{4 \pi^2} m^{\prime 2} \log \frac{4 \pi
f}{m^\prime}~. \eeq 
From the definition of $\lambda_t$ and
$\lambda_H$ one can show that $m^\prime
> \sqrt{2} f$, so with our bound on $f$ we have $m^\prime > 5.7$ TeV, which
for a 200 GeV Higgs implies a fine-tuning of 0.8\%.

The generic reason for obtaining relatively strong constraints on
the symmetry breaking scale in this model is that weak isospin is
violated. In the SM there is an $SU(2)$ global symmetry (called
``custodial'' $SU(2)$) which protects the $\rho_*$ parameter from
large corrections. These corrections in the SM can only come from
custodial $SU(2)$ violating interactions like hypercharge and
Yukawa couplings. The importance of custodial $SU(2)$ was noted in
the early literature on composite Higgs
models~\cite{KaplanGeorgi}, and recently emphasized again
in~\cite{Sekhar}. However, in the littlest Higgs model the heavy gauge
bosons break custodial symmetry since the embeddings of $SU(2)_1$ and $SU(2)_2$ into
$SU(5)$ as $\mathbf{5} \rightarrow (\mathbf{2},\mathbf{1}) + (\mathbf{1},\mathbf{2}) +
(\mathbf{1},\mathbf{1})$ break all non-Abelian global symmetries acting on
the Higgs.  Interactions of these heavy gauge bosons shift observables
from their standard model values.  Indeed, we find that the
$\rho_*$ parameter in (\ref{rhostar}) gets corrections of order
$v^2/f^2$ independently of the mixing angles (even in the
absence of a triplet VEV), therefore it is not hard to understand the
bounds found above.

\section{Conclusions}
\label{conclusions-sec}
\setcounter{equation}{0}
\setcounter{footnote}{0}

We have calculated the electroweak precision constraints on the
littlest Higgs model, incorporating corrections resulting from
heavy gauge boson exchange and the triplet VEV.  Using a global
fit to 21 observables, we found that generically throughout the
parameter space the smallest symmetry breaking scale consistent
with present experimental measurements is well above 4 TeV and for
particular parameters the bound is $f > 4.0$ TeV at 95\% C.L\@,
which implies that the Higgs mass squared is tuned to 0.8\%. This
bound arises for a specific choice of the high energy gauge
couplings, roughly $a=0.025$, $c = g/g_2 \sim 0.99$, and $c' =
g'/g'_2 \sim 0.3$. The origin of the strong constraints on this
model is the absence of a custodial $SU(2)$ symmetry, leading to
large contributions to $\Delta \rho_* = \alpha T$, even in the
absence of a triplet Higgs VEV. To the best of our knowledge, no
little Higgs model constructed to date has a custodial $SU(2)$
symmetry, suggesting that similarly strong constraints are
expected in other little Higgs models.

\section*{Acknowledgments}

G.D.K. and J.T. thank the particle theory group at Cornell
University for a very pleasant visit where this work was
completed. We thank Nima Arkani-Hamed, Sekhar Chivukula,
Ann Nelson, Martin Schmaltz, Elizabeth Simmons, and
Witold Skiba for helpful discussions and the Aspen Center for
Physics where this project originated. We also thank Nima Arkani-Hamed,
Andy Cohen, Bob McElrath and Ann Nelson for useful comments and suggestions
on the first version of this paper. The research of C.C., J.H.,
and P.M. is supported in part by the NSF under grant PHY-0139738,
and in part by the DOE OJI grant DE-FG02-01ER41206. The research
of G.D.K. is supported by the US Department of Energy under
contract DE-FG02-95ER40896. The research of J.T. is supported by
the US Department of Energy under contract W-7405-ENG-36.

\section*{Appendix A:  Predictions for  Electroweak Observables}
\renewcommand{\theequation}{A.\arabic{equation}}
\setcounter{equation}{0}
\setcounter{footnote}{0}

In this appendix we give the predictions for the shifts in
the electroweak precision observables due to new tree-level physics beyond the
SM in the $SU(5)/SO(5)$ littlest Higgs model.
The electroweak observables depend on three parameters,
$c,c'$ and $\Delta$.  Using the results given in~\cite{Sformulas,Burgess}
we find the following results:

\begin{eqnarray}
\Gamma_Z &=& \left( \Gamma_Z \right)_{SM} \left[1 +\Delta (1.7-0.23 c^2-0.89 c^4-3.8 c'^2+0.83 c'^4) -5.4 \Delta' \right]
 \nonumber \\
R_e &=& \left( R_e \right)_{SM} \left[1 +\Delta (0.34 -0.18 c^2+0.08 c^4-3.0 c'^2+4.6 c'^4) -1.1 \Delta' \right]
 \nonumber \\
R_\mu &=& \left( R_\mu \right)_{SM} \left[1 +\Delta (0.34 -0.18 c^2+0.08 c^4-3.0c'^2+4.6 c'^4) -1.1 \Delta' \right]
 \nonumber \\
R_\tau &=& \left( R_\tau \right)_{SM} \left[1 +\Delta (0.34 -0.18 c^2+0.08 c^4-3.0c'^2+4.6 c'^4) -1.1 \Delta' \right]
 \nonumber \\
\sigma_h &=& \left( \sigma_h \right)_{SM} \left[1 +\Delta (-0.04 +0.02 c^2+0.01 c^4+0.31 c'^2-0.48 c'^4) +0.12 \Delta' \right]
 \nonumber \\
R_b &=& \left( R_b \right)_{SM} \left[ 1 +\Delta (-0.08+0.04 c^2-0.02 c^4+0.66 c'^2-1.0 c'^4) +0.24 \Delta' \right]
 \nonumber \\
R_c &=& \left( R_c \right)_{SM} \left[1 +\Delta (0.15-0.08 c^2+0.04 c^4
-1.3 c'^2+1.9 c'^4) -0.47 \Delta' \right]
 \nonumber \\
A_{FB}^e &=& \left( A_{FB}^e \right)_{SM}  +\Delta(0.73-0.38 c^2 +0.18 c^4
-6.4 c'^2+9.8 c'^4) -2.3 \Delta' \nonumber \\
A_{FB}^\mu &=& \left( A_{FB}^\mu \right)_{SM}  +\Delta(0.73-0.38 c^2 +0.18 c^4
-6.4 c'^2+9.8 c'^4)-2.3 \Delta' \nonumber \\
A_{FB}^\tau &=& \left( A_{FB}^\tau \right)_{SM}+\Delta(0.73-0.38 c^2 +0.18 c^4
-6.4 c'^2+9.8 c'^4)-2.3 \Delta' \nonumber \\
A_{\tau}(P_\tau) &=& \left( A_{\tau}(P_\tau) \right)_{SM} +\Delta (3.2 -1.7 c^2+0.78 c^4-28 c'^2+43 c'^4) -10 \Delta'
\nonumber \\
A_{e}(P_\tau) &=& \left( A_{e}(P_\tau) \right)_{SM} +\Delta (3.2 -1.7 c^2+0.78 c^4-28 c'^2+43 c'^4) -10 \Delta'
\nonumber \\
A_{FB}^b &=& \left( A_{FB}^b \right)_{SM} +\Delta (2.3 -1.2 c^2+0.54 c^4-20 c'^2+30 c'^4) -7.2 \Delta' \nonumber \\
A_{FB}^c &=& \left( A_{FB}^c \right)_{SM} +\Delta (1.8 -0.91 c^2+0.42 c^4
-15 c'^2+23 c'^4) -5.6 \Delta' \nonumber \\
A_{LR} &=& \left( A_{LR} \right)_{SM} +\Delta (3.2 -1.7 c^2+0.78 c^4-28 c'^2+43 c'^4) -10 \Delta' \nonumber \\
M_W &=& \left( M_W \right)_{SM} \left[1+\Delta(0.89-0.21c^2+0.21c^4-3.6 c'^2+3.6 c'^4)-2.9 \Delta' \right] \nonumber \\
g_L^2(\nu N \rightarrow \nu X) &=&
\left( g_L^2(\nu N \rightarrow \nu X) \right)_{SM} +
\Delta(1.1 -0.16 c^2  + 0.25 c^4  -2.7 c'^2  +
1.2 c'^4 ) \nonumber \\ && -3.4 \Delta'  \nonumber \\
g_R^2(\nu N \rightarrow \nu X) &=&
\left( g_R^2(\nu N \rightarrow \nu X) \right)_{SM}
+ \Delta (-0.032 + 0.055 c^2  - 0.085 c^4  +
0.92 c'^2  \nonumber \\ && - 0.42 c'^4 ) +0.10 \Delta'  \nonumber \\
g_{eV}(\nu e \rightarrow \nu e) &=&
\left( g_{eV}(\nu e \rightarrow \nu e) \right)_{SM}+
\Delta (-0.87 + 0.43 c^2  - 0.66 c^4  +7.1 c'^2  -
3.3 c'^4 )\nonumber \\ &&+2.8 \Delta'  \nonumber \\
g_{eA}(\nu e \rightarrow \nu e) &=&
\left( g_{eA}(\nu e \rightarrow \nu e) \right)_{SM} -\frac{5\Delta}{8}
+2 \Delta' \nonumber \\
Q_W(Cs) &=& \left( Q_W(Cs) \right)_{SM}+ \Delta
(-1.5 - 47 c^2  + 73 c^4
-786 c'^2  + 363 c'^4)+4.7 \Delta' \nonumber \\
\end{eqnarray}
We also give in Table (\ref{table})
the experimental data \cite{ErlerLang,LEPEWG}
and the SM predictions used for our fit.
\begin{table}[!htp]
\begin{center}
\begin{tabular}{|c|c|c|}\hline
Quantity & Experiment & SM($m_h=115$ GeV) \\ \hline
$\Gamma_Z$ & 2.4952 $\pm$ 0.0023 & 2.4965 \\
$R_e$ & 20.804 $\pm$ 0.050 & 20.744 \\
$R_\mu$ & 20.785 $\pm$ 0.033 & 20.744 \\
$R_\tau$ & 20.764 $\pm$ 0.045 & 20.744 \\
$\sigma_h$ & 41.541 $\pm$ 0.037 & 41.480 \\
$R_b$ & 0.2165 $\pm$ 0.00065 & 0.2157  \\
$R_c$ & 0.1719 $\pm$ 0.0031 & 0.1723 \\
$A_{FB}^e$ & 0.0145 $\pm$ 0.0025 & 0.0163 \\
$A_{FB}^\mu$ & 0.0169 $\pm$ 0.0013 & 0.0163 \\
$A_{FB}^\tau$ & 0.0188 $\pm$ 0.0017 & 0.0163 \\
$A_{\tau}(P_\tau)$ & 0.1439 $\pm$ 0.0043 & 0.1475 \\
$A_{e}(P_\tau)$ & 0.1498 $\pm$ 0.0048 & 0.1475 \\
$A_{FB}^b$ & 0.0994 $\pm$ 0.0017 & 0.1034 \\
$A_{FB}^c$ & 0.0685 $\pm$ 0.0034 & 0.0739 \\
$A_{LR}$ & 0.1513 $\pm$ 0.0021 & 0.1475 \\
$M_W$ & 80.450 $\pm$ 0.034 & 80.389 \\
$g_L^2(\nu N \rightarrow \nu X)$ & 0.3020 $\pm$ 0.0019 & 0.3039 \\
$g_R^2(\nu N \rightarrow \nu X)$ & 0.0315 $\pm$ 0.0016 & 0.0301 \\
$g_{eA}(\nu e \rightarrow \nu e)$ & -0.507 $\pm$ 0.014 & -0.5065 \\
$g_{eV}(\nu e \rightarrow \nu e)$ & -0.040 $\pm$ 0.015 & -0.0397 \\
$Q_W(Cs)$ & -72.65 $\pm$ 0.44 & -73.11 \\
\hline
\end{tabular}
\end{center}
\caption{The experimental results~\cite{ErlerLang,LEPEWG}
and the SM predictions for the various
electroweak precision observables used for the fit. The SM predictions
are for $m_h=115$ GeV and $\alpha_s=0.12$ and
calculated~\cite{Erler} using GAPP~\cite{GAPP}.}
\label{table}
\end{table}


\newpage
\begin{thebibliography}{99}

{\small

\bibitem{little1}
N.~Arkani-Hamed, A.~G.~Cohen and H.~Georgi,
Phys.\ Lett.\ B {\bf 513}, 232 (2001)
{\tt [hep-ph/0105239]}.


\bibitem{littlest}
N.~Arkani-Hamed, A.~G.~Cohen, E.~Katz and A.~E.~Nelson,
JHEP {\bf 0207}, 034 (2002)
{\tt [hep-ph/0206021]}.


\bibitem{littlestmoose}
N.~Arkani-Hamed, A.~G.~Cohen, E.~Katz, A.~E.~Nelson, T.~Gregoire and J.~G.~Wacker,
JHEP {\bf 0208}, 021 (2002)
{\tt [hep-ph/0206020]}.

\bibitem{Witek}
I.~Low, W.~Skiba and D.~Smith,
Phys.\ Rev.\ D {\bf 66}, 072001 (2002)
{\tt [hep-ph/0207243]}.

\bibitem{littlepheno}
N.~Arkani-Hamed, A.~G.~Cohen, T.~Gregoire and J.~G.~Wacker,
JHEP {\bf 0208}, 020 (2002)
{\tt [hep-ph/0202089]};
K.~Lane,
Phys.\ Rev.\ D {\bf 65}, 115001 (2002)
{\tt [hep-ph/0202093]}.
T.~Gregoire and J.~G.~Wacker,
JHEP {\bf 0208}, 019 (2002)
{\tt [hep-ph/0206023]};
{\tt hep-ph/0207164}.

\bibitem{Sekhar}
R.~S.~Chivukula, N.~Evans and E.~H.~Simmons,
Phys.\ Rev.\ D {\bf 66}, 035008 (2002)
{\tt [hep-ph/0204193]}.

\bibitem{HiggsPseudo}
H.~Georgi and A.~Pais,
Phys.\ Rev.\ D {\bf 10}, 539 (1974);
Phys.\ Rev.\ D {\bf 12}, 508 (1975).

\bibitem{KaplanGeorgi}
D.~B.~Kaplan and H.~Georgi,
Phys.\ Lett.\ B {\bf 136}, 183 (1984);
D.~B.~Kaplan, H.~Georgi and S.~Dimopoulos,
Phys.\ Lett.\ B {\bf 136}, 187 (1984);
H.~Georgi, D.~B.~Kaplan and P.~Galison,
Phys.\ Lett.\ B {\bf 143}, 152 (1984);
M.~J.~Dugan, H.~Georgi and D.~B.~Kaplan,
Nucl.\ Phys.\ B {\bf 254}, 299 (1985).

\bibitem{KaplanGeorgiSU2}
D.~B.~Kaplan and H.~Georgi,
Phys.\ Lett.\ B {\bf 145}, 216 (1984).

\bibitem{Martin}
M.~Schmaltz,
{\tt hep-ph/0210415}.

\bibitem{Manton}
N.~S.~Manton,
Nucl.\ Phys.\ B {\bf 158}, 141 (1979)
P.~Forg\'acs and N.~S.~Manton,
Commun.\ Math.\ Phys.\  {\bf 72}, 15 (1980);
%
S.~Randjbar-Daemi, A.~Salam and J.~Strathdee,
Nucl.\ Phys.\ B {\bf 214}, 491 (1983);
D.~Kapetanakis and G.~Zoupanos,
Phys.\ Rept.\  {\bf 219}, 1 (1992);
G.~R.~Dvali, S.~Randjbar-Daemi and R.~Tabbash,
Phys.\ Rev.\ D {\bf 65}, 064021 (2002)
[{\tt hep-ph/0102307}].

\bibitem{extra1}
Y.~Hosotani,
Phys.\ Lett.\ B {\bf 126}, 309 (1983);
Phys.\ Lett.\ B {\bf 129}, 193 (1983);
Annals Phys.\  {\bf 190}, 233 (1989).

\bibitem{extra2}
H.~Hatanaka, T.~Inami and C.~S.~Lim,
Mod.\ Phys.\ Lett.\ A {\bf 13}, 2601 (1998)
[{\tt hep-th/9805067}];
H.~Hatanaka,
Prog.\ Theor.\ Phys.\  {\bf 102}, 407 (1999)
[{\tt hep-th/9905100}];
M.~Kubo, C.~S.~Lim and H.~Yamashita,
{\tt hep-ph/0111327}.

\bibitem{extra3}
I.~Antoniadis and K.~Benakli,
Phys.\ Lett.\ B {\bf 326}, 69 (1994)
[{\tt hep-th/9310151}];
I.~Antoniadis, K.~Benakli and M.~Quir\'os,
Nucl.\ Phys.\ B {\bf 583}, 35 (2000)
[{\tt hep-ph/0004091}];
I.~Antoniadis, K.~Benakli and M.~Quir\'os,
New J.\ Phys.\  {\bf 3}, 20 (2001)
[{\tt hep-th/0108005}];
G.~von Gersdorff, N.~Irges and M.~Quir\'os,
Nucl.\ Phys.\ B {\bf 635}, 127 (2002)
[{\tt hep-th/0204223}];
{\tt hep-ph/0206029};
{\tt hep-ph/0210134}.

\bibitem{extra4}
L.~J.~Hall, Y.~Nomura and D.~R.~Smith,
Nucl.\ Phys.\ B {\bf 639}, 307 (2002)
{\tt [hep-ph/0107331]}.
G.~Burdman and Y.~Nomura,
{\tt hep-ph/0210257}.

\bibitem{extra5}
C.~Cs\'aki, C.~Grojean and H.~Murayama,
{\tt hep-ph/0210133}.


\bibitem{Georgiununified}
H.~Georgi, E.~E.~Jenkins and E.~H.~Simmons,
Phys.\ Rev.\ Lett.\  {\bf 62}, 2789 (1989)
[Erratum-ibid.\  {\bf 63}, 1540 (1989)];
Nucl.\ Phys.\ B {\bf 331}, 541 (1990).


\bibitem{DimKap}
S.~Dimopoulos and D.~E.~Kaplan,
Phys.\ Lett.\ B {\bf 531}, 127 (2002)
{\tt [hep-ph/0201148]}.

\bibitem{CST}
R.~S.~Chivukula, E.~H.~Simmons and J.~Terning,
Phys.\ Lett.\ B {\bf 346}, 284 (1995)
{\tt [hep-ph/9412309]}.

\bibitem{CEKT}
C.~Cs\'aki, J.~Erlich, G.~D.~Kribs and J.~Terning,
Phys.\ Rev.\ D {\bf 66}, 075008 (2002)
{\tt [hep-ph/0204109]}.

\bibitem{Sformulas}
M.~E.~Peskin and T.~Takeuchi,
Phys.\ Rev.\ D {\bf 46}, 381 (1992).

\bibitem{ErlerLang}
J.~Erler and P.~Langacker, review in ``The Review of Particle Properties,''
K.~Hagiwara {\it et al.}  [Particle Data Group Collaboration],
Phys.\ Rev.\ D {\bf 66}, 010001 (2002),
updated version online: http://www-pdg.lbl.gov/2001/stanmodelrpp.ps.

\bibitem{Lynn}
D.~C.~Kennedy and B.~W.~Lynn,
Nucl.\ Phys.\ B {\bf 322}, 1 (1989);
D.~C.~Kennedy, B.~W.~Lynn, C.~J.~Im and R.~G.~Stuart,
Nucl.\ Phys.\ B {\bf 321}, 83 (1989).

\bibitem{Burgess}
C.~P.~Burgess, S.~Godfrey, H.~Konig, D.~London and I.~Maksymyk,
Phys.\ Rev.\ D {\bf 49}, 6115 (1994)
{\tt [hep-ph/9312291]}.

\bibitem{RSfit}
C.~Cs\'aki, J.~Erlich and J.~Terning,
Phys.\ Rev.\ D {\bf 66}, 064021 (2002)
{\tt [hep-ph/0203034]}.

\bibitem{LEPEWG}
LEP Electroweak Working Group, LEPEWWG/2002-01, \\
http://lepewwg.web.cern.ch/LEPEWWG/stanmod/.

\bibitem{Erler}
J.~Erler, private communication.

\bibitem{GAPP}
J.~Erler,
{\tt hep-ph/0005084}.

}
\end{thebibliography}
\end{document}